\documentclass[letterpaper]{JHEP3}
\usepackage[dvips]{graphicx}

\newcommand{\beq}{\begin{equation}}
\newcommand{\eeq}{\end{equation}}
\newcommand{\beqr}{\begin{displaymath}}
\newcommand{\eeqr}{\end{displaymath}}
\newcommand{\beqa}{\begin{eqnarray}}
\newcommand{\eeqa}{\end{eqnarray}}
\newcommand{\beqar}{\begin{eqnarray*}}
\newcommand{\eeqar}{\end{eqnarray*}}
\newcommand{\al}{\alpha}

\renewcommand{\b}{\beta}
\newcommand{\del}{\delta}

\newcommand{\eps}{\epsilon}
\newcommand{\gam}{\gamma}
\newcommand{\Gam}{\hat{\Gamma}}
\newcommand{\ga}{\gamma}
\newcommand{\G}{\Gamma}

\renewcommand{\l}{\lambda}

\newcommand{\veps}{\varepsilon}

\newcommand{\cF}{{\cal F}}
\newcommand{\cL}{{\cal L}}
\newcommand{\ssc}{\scriptscriptstyle}

\newcommand{\labell}[1]{\label{#1}} 
\newcommand{\reef}[1]{(\ref{#1})}

\newcommand{\non}{\nonumber}
\newcommand{\pf}{\partial}

\newcommand{\dir}[1]{\overline{#1}}

\newcommand{\tr}{\textrm{Tr}\;}

\newcommand{\df}{\textrm{d}}

\newcommand{\ph}{\phantom{1}}

\newcommand{\ie}{{\it i.e.,}\ }
\newcommand{\eg}{{\it e.g.,}\ }

\def\IR{{\hbox{{\rm I}\kern-.2em\hbox{\rm R}}}}

\newcommand{\ls}{\ell_s}

\newcommand{\T}{{\rm T}}
\newcommand{\bi}{\b_{\ssc I}}

\title{From D-$\dir{\textsf{D}}$ Pairs to Branes in Motion}

\author{Robert C. Myers\footnote{E-mail: {\tt rmyers@perimeterinstitute.ca}} ~and
David J. Winters\footnote{E-mail: {\tt winters@hep.physics.mcgill.ca}}\\
$^{*,\dagger}$ Perimeter Institute for Theoretical Physics,
Waterloo, Ontario N2J 2W9 Canada\\
$^{*,\dagger}$ Department of Physics, McGill University,
Montr\'eal,
Qu\' ebec H3A 2T8 Canada\\
$^{\ \ *}$ Department of Physics, University of Waterloo,
Waterloo, Ontario N2L 3G1 Canada}

\abstract{We investigate various supersymmetric brane
intersections. Motivated by the recent results on supertubes, we
investigate general constraints in which parallel brane-antibrane
configurations are supersymmetric. Dual descriptions of these
configurations involve systems of intersecting branes in relative
motion. In particular, we find new supersymmetric configurations
which are {\it not} related to a static brane intersection by a
boost. In these new configurations, the intersection point moves
at the speed of light. These systems provide interesting time
dependent backgrounds for open strings.}

 \preprint{hep-th/0211042}

\begin{document}

\section{Introduction}

D-branes\cite{joep} have played a central role in advancing our
understanding of string theory in the past seven years, and
despite the impressive collection of results that has accumulated
about these objects --- see, \eg ref.~\cite{clifford} --- they
continue to reveal new surprises. For example, Townsend and Mateos
\cite{stube} discovered a supersymmetric configuration in which a
cylindrical D2-brane is supported against collapse by the angular
momentum of crossed electric and magnetic fields on its
worldvolume. As a result of the worldvolume gauge fields, these
supertubes also lend themselves to an interpretation as bound
states of a cylindrical D2-brane with D0-branes and F-strings. In
particular, the latter constituents seem to define the
supersymmetries preserved by the supertube. An essential
ingredient in this result seems to be that the electric field
takes the `critical' value
\beq \labell{crite}
E=\pm{1\over2\pi l_s^2}\ .
\eeq
While not producing a vanishing brane tension, this critical field
further allows the cross-section of the cylinder to take an
arbitrary profile \cite{stube2}. In any event, as the
cross-section is closed, the system carries no overall D2 charge.
Roughly antipodal elements of the cylinder have the opposite
orientation and so can be considered as D2-$\dir{\textrm{D2}}$
pairs for which the charge cancels. From this point of view, the
fact that the configuration is supersymmetric comes as a surprise.

Ref.~\cite{ell} demonstrated, in the matrix theory description of
the supertube \cite{mattub}, that a one-quarter supersymmetric,
parallel D2-$\dir{\textrm{D2}}$ pair could be obtained as the
limit of a supertube of elliptical cross-section, in which the
major axis is taken to be infinitely large while the minor axis is
held fixed. In order to preserve charge, the tube necessarily
degenerates into a brane-antibrane pair. Again such a
configuration would typically would break all background
supersymmetry, however, as for the supertube, partial
supersymmetry is retained because of the presence of worldvolume
gauge field fluxes. Ref.~\cite{ell} discusses the supersymmetry
and stability of these configurations in the Born-Infeld,
supergravity and matrix theory descriptions, while in
ref.~\cite{strings}, similar issues are analysed from the
worldsheet perspective of open strings connecting the branes. It
is shown that the tachyonic mode which might naively appear at
small brane separations is lifted from the spectrum when the
supersymmetry conditions are satisfied. A similar discussion
appears in ref.~\cite{stube2}, while ref.~\cite{ddbar} provides a
similar analysis for D3-$\dir{\textrm{D3}}$ pairs. The electric
fields considered in refs.~\cite{stube2,ell,strings,ddbar} are
chosen to take the critical value \reef{crite}, which is natural,
given that their motivation comes from supertube physics.

In section \ref{pear}, we examine the general supersymmetry
conditions for parallel D2-D2 and D2-$\dir{\textrm{D2}}$ pairs
with gauge field fluxes. More precisely, we consider parallel
brane configurations in flat space and find a three parameter
family of worldvolume gauge fields that preserve a quarter of the
thirty-two background supersymmetries. In particular, we show that
a `critical' electric field \reef{crite} is not a necessary
condition. A similar result is implicit in the analysis of
D3-$\dir{\textrm{D3}}$ and D4-$\dir{\textrm{D4}}$ pairs found in
ref.~\cite{lug}.

In section \ref{duel}, we investigate several dual descriptions of
the above configurations. The lift to eleven dimensional M-theory
produces a pair of M2-branes oriented at an angle to one another
and also in
relative motion. Similarly, T-duality along the direction of the
electric fields produces angled D-strings in relative motion. In
both of these cases, the surprise is that supersymmetry is
preserved even though the branes are moving relative to one
another.

Hence, in section \ref{lorentz}, we provide a more general
discussion of branes in motion. An important ingredient to
producing a supersymmetric configuration is that the intersection
point travels at the speed of light. We note that such null
intersections were considered in ref.~\cite{bob2} in an M-theory
framework. We investigate these configurations from different
physical points of view, and find that although they are
supersymmetric, their perturbations seem unstable. We find that
the open string excitations connecting the branes cannot typically
follow the intersection point and so are stretched as they fall
behind.

The main text closes with a brief discussion of our results in
section \ref{disc}. Appendix \ref{hoho} presents some technical
details involved in the T-duality transformations appearing in
section \ref{duel}.

\section{D-$\dir{\textrm{D}}$ Pairs}
\label{pear}

The primary focus of this section is to examine general
constraints that arise from demanding that a parallel pair
consisting of a D-brane and an anti-D-brane are supersymmetric.
While naively such a configuration breaks all of the
supersymmetries, the system can be restored to being one-quarter
BPS by introducing fluxes of the worldvolume gauge fields
\cite{stube2,ell}. As in ref.~\cite{lug}, and in contrast to most
previous analyses, our generalized approach shows that the
supersymmetric configurations need not involve `critical' electric
fields \reef{crite}. The latter result could have been anticipated
by noting that the condition in eq.~\reef{crite} is not invariant
under Lorentz boosts in the worldvolume.

The following calculations focus on the supersymmetry conditions
for parallel D2-D2 and D2-$\dir{\textrm{D2}}$ pairs with gauge
field fluxes. We also restrict ourselves to cases in which the
electric fields on each brane can be Lorentz-transformed to lie in
the same direction.

\subsection{SUSY Pairs}

As described in ref.~\cite{BKOP}, the condition for a D-brane to
preserve some of the supersymmetry of its background is based upon
the invariance of the worldvolume action under local
$\kappa$-symmetry and global spacetime supersymmetry
transformations of the spacetime fermion fields. Making a
combination of these transformations, and gauge fixing the
$\kappa$-symmetry, results in a global worldvolume supersymmetry.
By requiring consistency of the gauge choice under further
supersymmetry transformations, one finds a constraint on the
Killing spinors, $\eps$, that generate the supersymmetries of the
background. This demands that the generators satisfy
$\G\eps=\pm\eps$, where $\G$ is a Hermitian traceless product
structure (\ie $\tr\G=0$ and $\G^2=1$) appearing in the
$\kappa$-symmetry transformation. The sign on the right-hand side
of this condition is taken to be positive for a brane and negative for an antibrane.
For a D2- or $\dir{\textrm{D2}}$-brane, the product structure is
given by
\beq
\G=\frac{1}{\sqrt{-\det(g+\cF)}}\left(1+\frac{1}{2}\Gam^{ab}\cF_{ab}
\G_{11}\right)\Gam_{012}\ , \labell{proj} \eeq
where $\cF=B+2\pi\ls^2 F$ is the generalised Born-Infeld (BI)
field strength and $\Gam_a$ denote the `pull-backs' of the
spacetime Dirac matrices to the worldvolume --- see
eq.~\reef{weegam} below. If we choose a background coordinate
system, $x^\mu$, then the induced metric on the worldvolume can be
written in terms of the ten-dimensional \textit{zehnbein}
$E^{\underline{\al}}_{\ph\mu}$, as
\beq
g_{ab}=\pf_a x^\mu \pf_b x^\nu G_{\mu\nu}=(\pf_a x^\mu
E^{\underline{\al}}_{\ph\mu}) (\pf_b x^\nu
E^{\underline{\beta}}_{\ph\nu})\eta_{\underline{\al\beta}}\ ,
\eeq
Then, the $\Gam_a$ are defined as
\beq
\Gam_a=(\pf_a x^\mu E^{\underline{\al}}_{\ph\mu})
\G_{\underline{\al}}\ , \label{weegam}
\eeq
where $\G_{\underline{\al}}$ are the constant, ten-dimensional
Dirac matrices with
$\{\G_{\underline{\al}},\G_{\underline{\beta}}\}=2
\eta_{\underline{\al\beta}}$. By $\G_{11}=\G_0\cdots\G_9$ we
denote the ten-dimensional chirality operator. In the analysis
that follows we begin by considering a pair of parallel D2-branes
carrying fluxes. The extension to a D2-$\dir{\textrm{D2}}$ pair
appears at the end of the section --- all pertinent results are
presented, but the details omitted since the analyses are so
similar.

We work exclusively in a Minkowski space background. We adopt a
Cartesian coordinate frame, so the background Killing spinors
$\eps$ are constant and arbitrary (up to being non-zero). Also,
the \textit{zehnbein} is trivially
$E^{\underline{\al}}_{\ph\mu}=\del^{\underline{\al}}_{\ph\mu}$ so
that $\Gam_a=(\pf_a x^\mu) \G_\mu$. Hence, we see that for a flat
D2-brane aligned, in static gauge, with the $t$, $x$ and $y$
directions of this background, we have $\Gam_0=\Gam_t=\G_t$,
$\Gam_1=\Gam_x=\G_x$, $\Gam_2=\Gam_y=\G_y$ and that
eq.~\reef{proj} reduces to
\beq
\G=\frac{1}{\sqrt{-\det(\eta+\cF)}}\left(\G_{txy}+
(\G_y\cF_{tx}+\G_x\cF_{yt}+\G_t\cF_{xy})\G_{11}\right). \label{flatcond}
\eeq
Due to the properties of $\G$, the single condition $\G\eps=\eps$
reduces  the number of independent components of $\eps$ by a
factor of two. That is, a single D2-brane with arbitrary worldvolume
fluxes is one-half BPS. For a multi-brane configuration, one
tests the mutual compatibility of the single-brane conditions. In
general, this reduces the number of independent components of the
surviving Killing spinor $\eps$ to $k$, say, in which case the
fraction of supersymmetry preserved is $k/32$. Incompatibility of
the single-brane conditions implies $k=0$, meaning that
supersymmetry is completely broken.

We consider a pair of flat, parallel D2-branes parametrised with
the  background coordinates $(t,x,y)$, as above. They may be
separated by a finite distance in the transverse dimensions. In
addition, they have constant BI field strengths of the form
\beq F_i=E_i\, \df y \wedge \df t + B_i\, \df y \wedge \df x\ ,
\qquad (i=1,2) \label{genF} \eeq
while we assume that the Kalb-Ramond field vanishes. This ansatz
incorporates many possible field configurations. In particular,
any configuration in which at least one of the branes, say the
second, has $F_2\cdot F_2\ne 0$ is Lorentz-equivalent to one of
the form \reef{genF} with either $E_2=0$ (if $F_2\cdot F_2>0$) or
$B_2=0$ (if $F_2\cdot F_2<0$). In many cases it is also possible
to boost to a frame in which either $E_1$ or $E_2$ takes the
critical value, but we will not restrict ourselves in this way.

Given this ansatz for the fields, the supersymmetry conditions
become, from eq.~\reef{flatcond},
\beq
\G_i\eps=\frac{1}{\cL_i}(\G_{txy}+2\pi\ls^2\, E_i\G_x\G_{11}-
2\pi\ls^2\, B_i\G_t\G_{11})\eps=\eps\ , \qquad (i=1,2)
\label{Gipar}
\eeq
where $\cL_i=\sqrt{-\det(\eta+F_i)}=\sqrt{1-4\pi^2\ls^4\,
E_i^2+4\pi^2\ls^4\, B_i^2}$. The commutator of $\G_1$ and $\G_2$
provides the useful result
\beq
[\G_1,\G_2]\eps=0 \quad\Leftrightarrow\quad
\left((E_1-E_2)\G_{ty}\G_{11}-(B_1-B_2)\G_{xy}\G_{11}+2\pi\ls^2
(E_1B_2-E_2 B_1)\G_{xt}\right)\eps=0\ .
\label{parcomm}
\eeq
Using eqs.~\reef{Gipar} and \reef{parcomm}, and the properties of
the gamma operators, we find that
\beq
\G_y\eps=\frac{4\pi^2\ls^4 (E_1 B_2-E_2B_1)}{\cL_2-\cL_1}\eps
             =\frac{B_2\cL_1- B_1\cL_2}{E_1-E_2}\eps
             =\frac{E_2\cL_1- E_1\cL_2}{B_1-B_2}\eps\ . \label{Gzcond}
\eeq
Further, we note that the eigenvalues of $\G_y$ are $\pm1$, as
$\G^2_y=1$. Therefore the above seems to provide three constraints
on the four independent field strengths. Remarkably, however, one
finds that there remains a three parameter family of solutions, as
follows. For each eigenvalue of $\G_y$, eq.~\reef{Gzcond} provides
three expressions for, say, $E_1$. Each of these is a solution to
a quadratic equation and so contains an arbitrary sign. For a
suitable choice of these signs, the three expressions are
identical. This gives the most general solution for $E_1$, as a
function of the other three fields. Hence, we have the auxiliary
projection
\beq
\G_y\eps=\pm\eps\ ,\label{Gzpar}
\eeq
which applies with the matching field configuration
\beq E_1=E_1^\pm\equiv\frac{E_2(1+4\pi^2\ls^4
B_1B_2)\pm(B_2-B_1)\cL_2}{1+4\pi^2\ls^4 B_2^2}\ . \labell{parsol}
\eeq
Here one simultaneously chooses either the plus or the minus sign
in each of these equations. These two expressions follow from
eqs.~\reef{Gipar} and so are necessary for supersymmetry, but they
may not be sufficient to ensure it. To check this we must find
under what additional conditions, if any, the imposition of
eqs.~\reef{Gzpar} and \reef{parsol} guarantees the compatibility
of eqs.~\reef{Gipar}. By considering each supersymmetry condition
in turn we find that they are actually equivalent, and not merely
compatible, if
\beq \G_y\eps=\pm\eps\ ,\quad E_1=E_1^\pm\ ,\quad\textrm{and}\quad
(1+4\pi^2\ls^4\,B_1B_2)\cL_2\pm4\pi^2\ls^4\,E_2(B_1-B_2)>0\ .
\labell{fullsol} \eeq

Hence, the amount of supersymmetry preserved by these branes is
determined by the compatibility of the operators $\G_y$ and
$\G_1$, say, given that we choose a valid field configuration from
the region of parameter space allowed by the inequality. We note
that, like $\G_1$,  $\G_y$ is traceless and squares to the
identity. Furthermore, it can easily be checked that the two operators
commute. By solving the two conditions simultaneously we find that
$1/4$ of the background supersymmetry is preserved. What is more,
the simple form of the auxiliary projection is virtually
independent of the brane configuration (the only dependence being
that it projects into the volume of the brane). This means that
the same conditions can be applied to find supersymmetric
configurations of any number of parallel branes, so long as each
of the field strengths takes the form \reef{genF} and is related
to that on an arbitrarily chosen reference brane by
eq.~\reef{parsol}.

This entire calculation can be easily modified to consider the
case  of a parallel D2-$\dir{\textrm{D2}}$ pair. In the heuristic
picture of an anti-D2 brane as an `upside-down' D2-brane, we see
that the transition from a D2 to an anti-D2 involves a parity
transformation of one of the worldvolume coordinates, in which
case $\G\to -\G$. As advertised above, the supersymmetry condition
for a flat anti-D2 brane is therefore $\G\eps=-\eps$, with $\G$ as
in eq.~\reef{flatcond}. Assuming an ansatz of the form
\reef{genF}, with $F_2$ on the antibrane, the supersymmetry
conditions become
\beq \G_i \eps=\frac{1}{\cL_i}(\G_{txy}+2\pi\ls^2\,
E_i\G_x\G_{11}-2\pi\ls^2\, B_i\G_t\G_{11})\eps=s_i \eps\ , \qquad
(i=1,2) \label{bparcond} \eeq
where $s_1=1$ and $s_2=-1$. A repetition of the above calculation
shows that one-quarter supersymmetric, parallel D2-$\dir{\textrm{D2}}$
pairs exist under the conditions
\beqa \G_y\eps=\pm\eps\ ,&&\quad
E_1=\widetilde{E}^\pm_1\equiv\frac{E_2(1+4\pi^2\ls^4\, B_1B_2)
\mp(B_2-B_1)\cL_2}{1+4\pi^2\ls^4\, B_2^2} \non\\
\textrm{and}&&\quad (1+4\pi^2\ls^4\, B_1 B_2)\cL_2\mp
4\pi^2\ls^4\, E_2(B_1-B_2)<0\ . \label{babconf} \eeqa
Note the subtle difference in signs between these expressions and
those appearing in eq.~\reef{fullsol}. Just as in the D2-D2 case,
these results imply the equivalence of the single-brane conditions
and so can be used to establish the supersymmetry of a collection
of arbitrarily many branes and antibranes by relating the gauge
fields on each (anti)brane to those on a reference brane using the
appropriate expression, eq.~\reef{parsol} or eq.~\reef{babconf}. A
similar configuration of branes and antibranes was studied in
ref.~\cite{ell}, in which all the electric fields were critical.
In that case the system explicitly preserves the supersymmetries
corresponding to the presence of F-strings and D0-branes. That is,
the Killing spinors satisfy
\beq \labell{karchproj}
\textrm{F1:}\quad\G_{ty}\eps=-\G_{11}\eps\qquad\textrm{and}\qquad
\textrm{D0:}\quad\G_t\G_{11}\eps=\eps\ .
\eeq
In our conventions, the supersymmetric field choices of ref.~\cite{ell} are
\beq
2\pi\ls^2 E_1=-1\ , \quad B_1<0\ , \quad 2\pi\ls^2 E_2=-1 \quad
\textrm{and} \quad B_2>0\ .
\eeq
Substituting the latter three into
the conditions \reef{babconf} as parameters, one finds the
$\G_y\eps=\eps$ solution $2\pi\ls^2 E_1=-1$, so these choices are
consistent with our analysis. The auxiliary projection emerges
trivially from manipulating the F-string and D0-brane
supersymmetry projections \reef{karchproj}.

\subsection{Dual Descriptions}
\label{duel}

We now describe various dual interpretations of the
brane configurations discussed above. These are
gained by the usual methods of T-duality, and by lifting to
M-theory.

Consider the transition from ten to eleven dimensions.
The gauge degrees of freedom on a D-brane are traded for
geometrical degrees of freedom describing the orientation and
velocity of the dual M2-brane in the eleventh dimension. The
details of this interchange, made explicit in the formal
derivation of the M2-brane action from the D2-brane
action\cite{deAS,christ}, are (assuming a flat space background)
\beq \labell{prof} \pf_a
z=\frac{\pi\ls^2\,\eps_{abc}F^{bc}}{\sqrt{1+2\pi^2\ls^4\,F^2}}\ ,
\eeq
where $z$ is the eleventh dimension and $F$ is the BI field
strength.\footnote{In general, the pull-back of the RR one-form
also enters this expression but we have set it to zero.} The
partial derivatives are with respect to the worldvolume
coordinates on the D2-brane.

To illustrate this, we consider a supersymmetric pair of D2-branes
with worldvolume gauge fields
\beq E_1= B_1=0\ , \quad E_2=\pm B_2=-\varepsilon\ .
\label{specsol} \eeq
One may verify that this choice is compatible with
eq.~\reef{fullsol}. According to eq.~\reef{prof}, the D2-brane
carrying the fluxes lifts to an M2-brane with
\beq \pf_x z=\pm\pf_t z=2\pi\ls^2 \veps\ , \label{susyprof1} \eeq
while the other simply becomes to a flat, stationary M2-brane with
$z=0$. Hence, one of the M2-branes is rotated and boosted into the
eleventh dimension, while the other remains at rest in the $(x,y)$
plane. Note that the branes may not intersect, as they may be
displaced in the overall transverse dimensions. Supersymmetry is
preserved in the M-theory lift, so these M2-branes remain
$1/4$-supersymmetric. This is an interesting result, as it is well
known that, generically, branes at angles do not preserve
supersymmetry. Exceptional classes of supersymmetric branes exist
\cite{BDL,OT,bob1}, however, they would involve rotating the
branes simultaneously in at least two orthogonal planes --- this
need not be true for bound states, as we will see below. In
contrast, the second M2-brane above is only rotated in the $(x,z)$
plane relative to the first and so the additional motion of this
brane must be essential to preserving supersymmetry. We note that
the speed with which the intersection point between the two branes
travels (or rather the speed of the intersection of the moving
M2-brane with a constant $z$ hypersurface) is easily determined
from eq.~\reef{susyprof1} to be
\beq \labell{foreshadow}
\left.{dx\over dt}\right|_I={\pf_t z\over\pf_x z}=\pm1\ .
\eeq
As we will discuss in the next section, the fact that the brane
intersection moves at the speed of light is a necessary condition
for supersymmetry. Similar observations were made in classifying
the most general supersymmetric configurations of two M5-branes
\cite{bob2}.

Note that if we choose $B_1=B_2$, the only solution for the
electric fields is $E_1=E_2$. In lifting these to M-theory they
are therefore rotated by the same angle and become parallel
M2-branes. This indicates more generally that non-parallel
M2-branes can
only preserve supersymmetry if they are in relative motion. A
final comment is that, clearly, by returning to ten dimensions by
compactifying a direction transverse to all of the worldvolume
directions, one recovers identical configurations of D2-branes
with vanishing gauge fields.

Other related brane configurations can be found using T-duality.
Of course, dualising in the transverse directions will simply
produce higher-dimensional versions of those described above. More
interesting are the T-dualities along the directions parallel to
the branes, which we discuss here.

A D2-brane with a general BI field strength is interpreted as a
D2-D0-F1 bound state and T-dualising parallel to its worldvolume
produces a D1-F1 bound state with momentum in the plane of the
initial system (see the appendix for details). This observation
allows one to relate the supertube configurations to a super-helix
\cite{shelix,scurve} via a T-duality along the axis of the
cylinder. A supersymmetric pair of parallel D2-branes, as
discussed above, will therefore have a (supersymmetric) dual that
consists of two D-strings in relative motion. In general the
D-strings will \textit{not} be parallel, so here we again see the
preservation of supersymmetry in an unusual setting, involved with
the motion of the branes. A more detailed discussion of these
configurations will be postponed to section \ref{lorentz}.

For our D2-D2 solutions in section 2.1, momentum arises from
T-duality along the $y$ axis, and the D-strings carry no electric
flux, \ie they are pure D-strings, rather than D1-F1 bound states.
If we T-dualise along the $x$ axis, on the other hand, we find
static configurations of angled D1-F1 bound states, or
$(p,1)$-strings. This is in contrast to the situation in M-theory,
where relative motion is an essential ingredient in the
supersymmetry of non-parallel M-branes. As an example, consider
again the simple gauge field configuration \reef{specsol}. The
brane with vanishing gauge fields T-dualises to a stationary
$(0,1)$-string (\ie a D-string) parallel to the $y$ axis. The
other T-dualises to a $(p,1)$-string that is tilted in the ($x,y$)
plane. The tilt angle, $\theta$, with respect to the $x$ axis, and
the electric field, $e$, on the D-string, are related to $E_1$ and
$B_1$ by
\beq \labell{dualfield}
E_1=\frac{e}{\sin\theta}\ , \qquad B_1=\frac{\cot\theta}{2\pi\ls^2}\ .
\eeq
Therefore, the system of D-strings is supersymmetric when
\beq \labell{stringsol}
e=\pm\frac{\sin\phi}{2\pi\ls^2}\ ,
\eeq
expressed in terms of $\phi=\pi/2-\theta$, which is the angle
between the two branes. This is in agreement with the standard
result, $e=0$, for parallel D-strings, and implies that orthogonal
D-strings are supersymmetric for $|e|=(2\pi\ls^2)^{-1}$, \ie
critical electric flux. Since $e$ is the only component of the BI
field strength on the D-string, this would signal the vanishing of
its tension.

It is interesting to consider the case of intersecting D-strings,
\ie the configurations above where the D-strings are not displaced
in the overall transverse directions. In this case, we can relate
the results to discussions on string networks
\cite{net,junc,net2}.  In particular, the intersecting D-string
configuration above can be thought of as a special case of a two
vertex string network, in which the string joining the two
vertices has been shrunk to zero length --- see figure 1. When
each string of type $(p,q)$ has the same orientation within the
network as all the other strings of that type, the network is
$1/4$ supersymmetric \cite{net,junc,net2}. From the figure, it can
be seen that the two vertex network indeed satisfies this
condition. We have chosen the orientations and charges such that,
when the internal string is shrunk to zero length, the diagram
reduces to the D-string configuration discussed above.

To verify this interpretation, we shrink the connecting
$(p,0)$-string in figure 1 to recover the crossed D-strings
without changing $\theta$. The latter should then be related to
the flux on the $(p,1)$-string by the supersymmetry condition
\reef{stringsol}: $2\pi\ls^2|e|=|\cos\theta|$. Standard results
\cite{junc} on three string junctions allow the angle $\theta$ to
be expressed as
\beq \cos\theta=\frac{|p|g_s}{\sqrt{p^2g_s^2+1}}\ .
\label{thetajunk} \eeq
So our interpretation is valid if the electric flux on the
$(p,1)$-string is related to $p$ by
\beq \l |e|=\frac{|p|g_s}{\sqrt{p^2g_s^2+1}}\ .\label{pflux} \eeq
The latter is verified by a standard calculation relating the
number of F-strings to the electric displacement on the D-string
--- see, \eg ref.~\cite{core}. Moreover, the two configurations preserve
the same amount of supersymmetry, so we conclude that our
interpretation is correct.
\begin{center}
\includegraphics[scale=0.5]{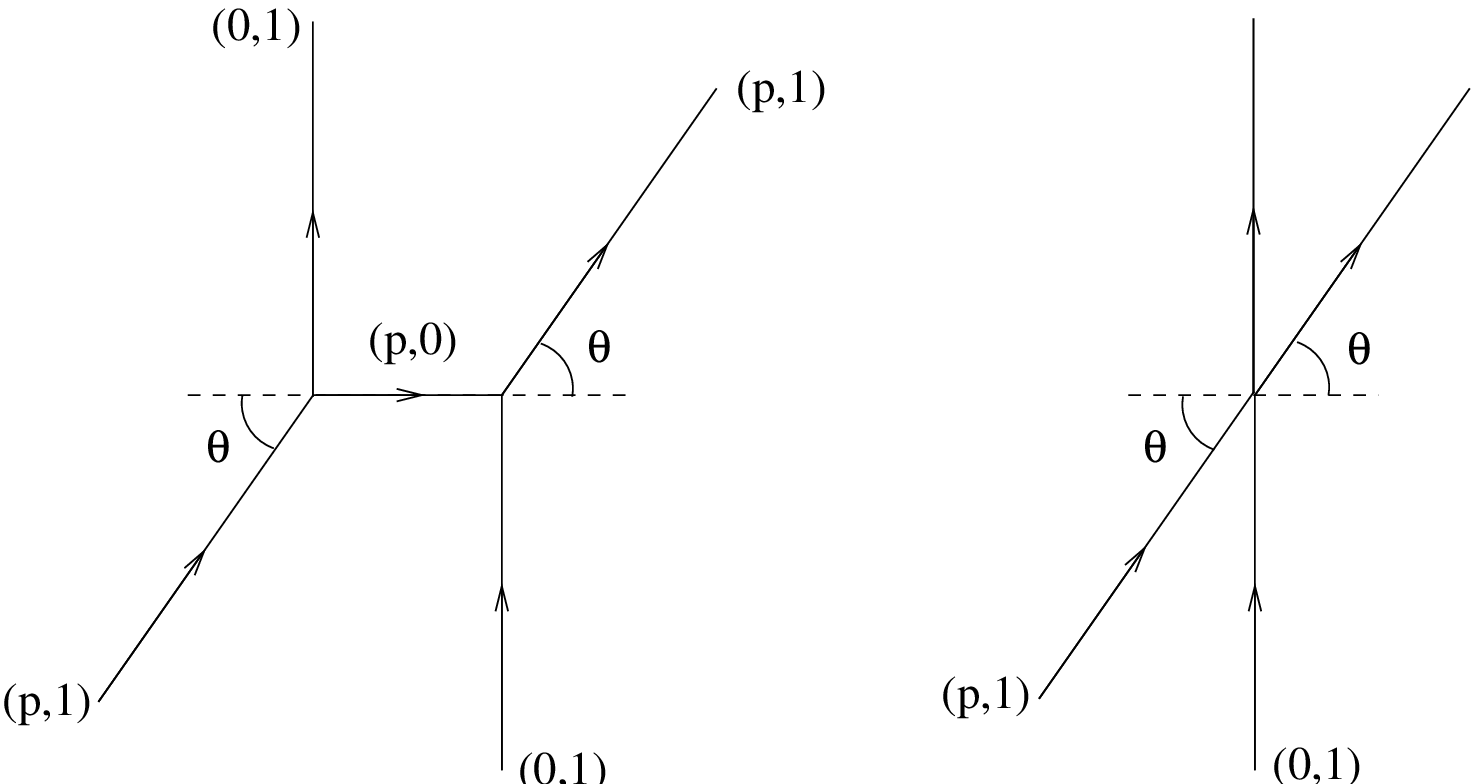}\\
Figure 1: The two vertex string network and the limiting case in\\
which the internal  $(p,0)$-string has been shrunk to zero length.
\end{center}

It was argued earlier that our results could also be used to find
supersymmetric stacks of more than two parallel D2-branes. The
above discussion of dual configurations can therefore be extended
to include an arbitrary number of intersecting D-strings. Of
course, T-dualities perpendicular to the plane of the D-strings
will produce supersymmetric systems of intersecting D$p$-branes. The
D$p$-branes will inherit an identical electric field from the
corresponding D-string. Hence the supersymmetry condition will be
the same as eq.~\reef{stringsol}.

\section{Branes in Motion}
\label{lorentz}

The presence of non-trivial gauge fields preserving supersymmetry
has been seen to produce interesting systems dual to the D2-D2 or
D2-$\dir{\textrm{D2}}$ pairs, both in string theory and M-theory.
In both cases the relevant duality generally mixes, or replaces,
the fluxes with angles and {\it momenta}. This motivates a
discussion in this section of more general configurations where
the constituent branes are in relative motion. It is useful to
consider the action of Lorentz boosts on D-branes, so we provide
the following discussion.

\subsection{Dirichlet meets Lorentz (and Pythagorus)}

The essential observation is that a(n unexcited) D-brane cannot
support any momentum in directions along its worldvolume. Hence
consider an observer moving at an angle to a D-brane, as
illustrated in figure 2(a). If one boosts to the frame of reference
of this observer (see figure 2(b)), one finds that the spatial
momentum vector describing the moving D-brane is orthogonal to the
angled brane. That is, the D-brane momentum is {\it not}
antiparallel to the observer's original velocity.

\begin{center}
\includegraphics[scale=0.6]{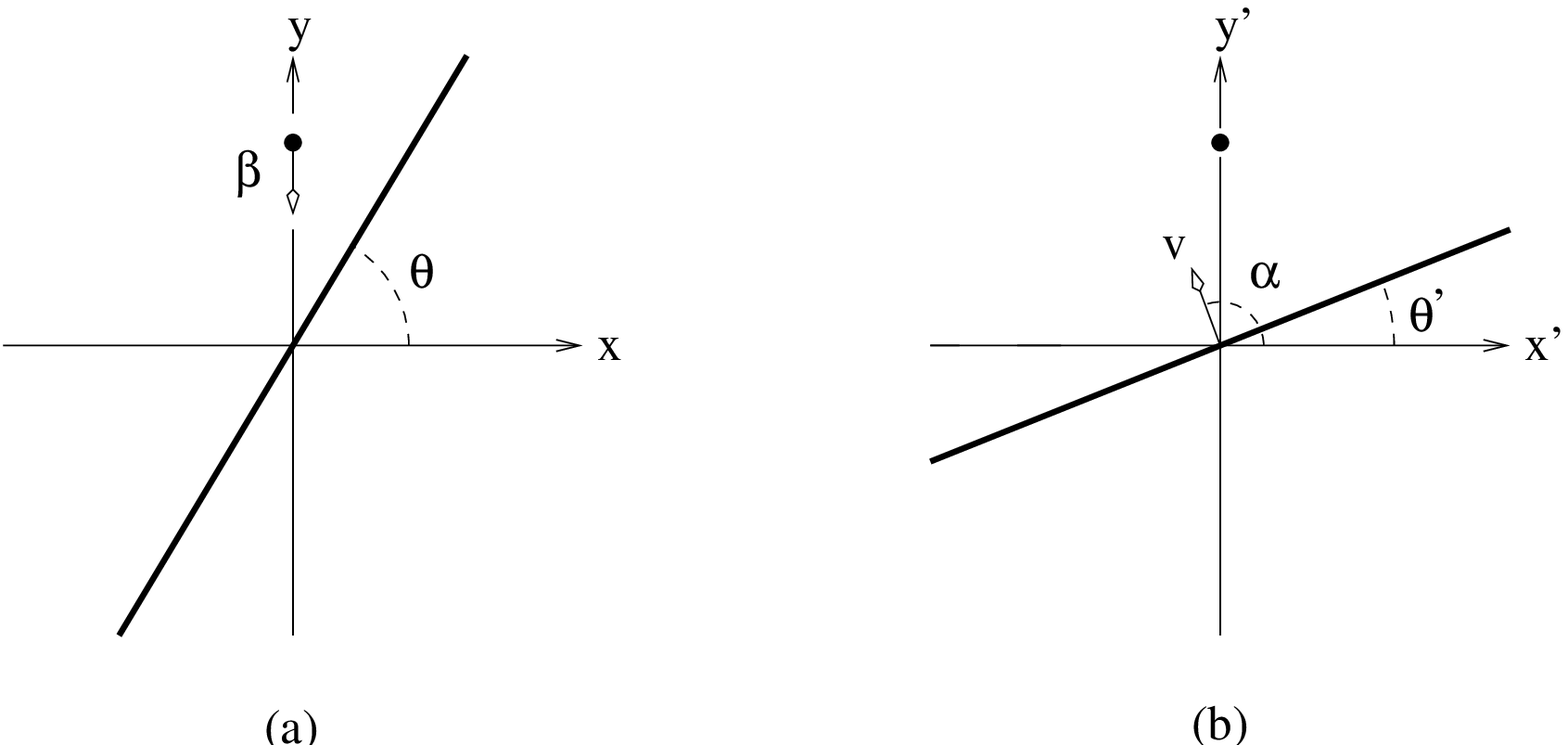}\\
Figure 2: Two perspectives on an angled brane
\end{center}

Of course, this amusing result arises because of the relativistic
nature of the D-brane's stress energy tensor. We can make precise
the qualitative observations above by considering the application
of a boost to the stress energy tensor. We will frame the
following discussion in the context of a D-string angled in the
($x,y$) plane, as shown in the figure. The same formulae obviously
apply for a higher dimensional D-brane extending off in orthogonal
directions, since only the components in ($t,x,y$) directions will
be relevant. Furthermore, we miss nothing by omitting the
$\delta$-functions which may localise the brane's stress-energy
tensor in some of these additional directions.

The stress energy tensor for a D-brane at rest at an angle
$\theta$ to the $x$ axis is easily derived to be
\beq \labell{stresstat} T_{ab}=\T\,\left[ \matrix{1& 0 & 0\cr
    0& -\cos^2\theta & -\cos\theta\sin\theta \cr
    0 & -\cos\theta\sin\theta  & -\sin^2\theta \cr}
\right]\,\delta(y\cos\theta-x\sin\theta)\ , \eeq
where T is the brane tension. It is easy to see that for
$\theta=0$ this expression reduces to a diagonal matrix,
$T_{ab}=\T\,diag(1,-1,0)\,\delta(y)$, as expected for a D-brane
lying along the $x$ axis. Now applying a boost along the $y$ axis
with velocity $\b$, so as to bring the observer in figure 1(a) to
rest, one finds
\beq \labell{stressmov} T_{a'b'}=\T\,\left[
\matrix{\ga^2(1-\b^2\sin^2\theta) & \b\ga\sin\theta\cos\theta &
-\b\ga^2\cos^2\theta \cr
   \b\ga\sin\theta\cos\theta & -\cos^2\theta & -\ga\sin\theta\cos\theta \cr
   -\b\ga^2\cos^2\theta & -\ga\sin\theta\cos\theta  & \ga^2(\b^2-\sin^2\theta) \cr}
\right]\,\delta(\ga(y'-\b t')\cos\theta-x'\sin\theta)\ , \eeq
where as usual $\ga=1/\sqrt{1-\b^2}$. Note that for
$\theta=\pi/2$, the stress energy reduces to
$T_{a'b'}=\T\,diag(1,0,-1)\,\delta(x')$. That is, this corresponds
to a D-brane lying on the $y$ axis and, since the boost is then in
the worldvolume of the (unexcited) brane, the configuration and
the stress energy tensor are left invariant. In general, from the
off-diagonal components $T_{t'i'}$ one sees that the boosted brane
has a momentum density with components along both the $x'$ and
$y'$ axes, even though the boost was along the $y'$ axis. The
angle of the motion with the $x'$ axis is given by
\beq \labell{anglep} \tan\alpha={T_{t'y'}\over
T_{t'x'}}=-\frac{\ga}{\tan\theta}\ , \eeq
while the angle which the brane makes with the $x'$ axis may be
read off from the argument of the $\delta$-function to be
\beq \labell{anglet} \tan\theta'=\frac{\tan\theta}{\ga}\ , \eeq
at any instant (in time $t'$). Hence we have
$\tan\alpha\,\tan\theta'=-1$, indicating that, as expected, the
spatial momentum of the boosted brane is orthogonal to its spatial
worldvolume. One observation about this geometry is that even
though the brane moves off at an angle $\alpha$ in the ($x'$,$y'$)
plane, the intersection point of the brane with the $y'$ axis
($x'=0$) moves with precisely the expected velocity $\beta$. One
final comment is that both the static and boosted branes will be
equally supersymmetric.

As mentioned above, this `spontaneous' generation of momentum
orthogonal to the boost direction applies for any extended
relativistic brane boosted at an angle to its worldvolume. This
includes D-branes, M-branes, fundamental strings and NS5-branes.
The effect manifests itself in interesting ways in configurations
dual to an angled brane. We briefly consider two examples here.

Consider `T-dualising' a D-string, angled as in figure 2(a), along
the $x$ axis.\footnote{Note that implicitly in the following we do
not consider the $x$ coordinate to be compact for either
configuration and so they are not strictly speaking T-dual to each
other.} As already implicit in the discussion of section 2.2, the
dual configuration corresponds to a D2-brane filling the ($x$,$y$)
plane and carrying a uniform magnetic flux
--- see, \eg ref.~\cite{clifford,more}. The latter flux may be
interpreted as constant density of D0-branes dissolved in the
worldvolume of the D2-brane:
\beq \labell{flux} \rho_{\textrm{\tiny D0}}={1\over
2\pi}F_{yx}={\cot\theta\over(2\pi\ls)^2}\ . \eeq
In this case, because of the worldvolume excitation, \ie the
magnetic field, the system is not invariant under boosts parallel
to the D2-brane. Rather, a boost in, say, $y$,  generates an
electric field
\beq \labell{fluxb} F_{x't'}=\ga\b{\cot\theta\over2\pi\ls^2}\ ,
\qquad\qquad F_{y'x'}=\ga{\cot\theta\over2\pi\ls^2}\ . \eeq
The appearance of the electric field corresponds to the
`spontaneous' generation of a density of fundamental strings
aligned along the $x'$ axis:
\beq \labell{fluxf} \rho_{\textrm{\tiny
F1}}=\frac{\ga\b\cos\theta}{2\pi\ls g_s}\ ,\eeq
corresponding to the electric displacement calculated from the
Born-Infeld action --- see, \eg ref.~\cite{core}. Note that in
this boosted configuration as well as winding number along the
$x'$ axis, there is momentum along the $y'$ axis since
$T_{t'y'}\propto F_{t'x'}F_{y'x'}$ is nonvanishing. The presence
of both of these is expected, in order to match the T-dual
configuration, \ie the boosted D-string carrying both $x'$ and
$y'$ components of momentum density. This is only possible because
the operations of Lorentz boost and T-duality commute if they are
performed at right-angles to one another. Of course, this subtle
interplay fails when both the T-duality and the boost are made in
the same direction. Then the two operations do not commute, as the
momentum generated by a boost in a certain direction is affected
by the subsequent action of T-duality in that direction. These
assertions may be verified explicitly by the reader, using the
T-duality relations given in the appendix.

Another interesting case considered in section 2.2 is to
reinterpret a D2-brane with fluxes in terms of its lift to
M-theory, as described by eq.~\reef{prof}. The D0-D2 bound state
above, with only a magnetic flux, corresponds to a M2-brane moving
in the eleventh dimension. Boosting along the $y$ axis introduces
an electric field parallel to the $x$ axis on the D2-brane, which
corresponds to tilting the M2-brane in the ($z$,$y$) plane.
(Recall that $z$ is the eleventh dimension.) Note that, just as
the magnetic field on the D2-brane is a worldvolume excitation and
so causes the D-brane to `feel' the boost, the motion of the
original M2-brane orthogonal to the boost axis can also be
regarded as a worldvolume excitation and, as a result, the system
is not invariant even though the boost is `along the brane' as
seen at any given instant of time.

Alternatively, one could consider beginning with only an electric
flux $F_{xt}$, \ie a D2-F1 bound state, in which case a boost in
$y$ introduces a magnetic flux $F_{yx}$. The M-theory description
is then analogous to the description of boosting a tilted D-brane
given above. One begins with an M2-brane tilted in the ($z$,$y$)
plane and boosting along the $y$ axis spontaneously generates a
component of momentum in the $z$ direction.

\begin{center}
\includegraphics[scale=0.6]{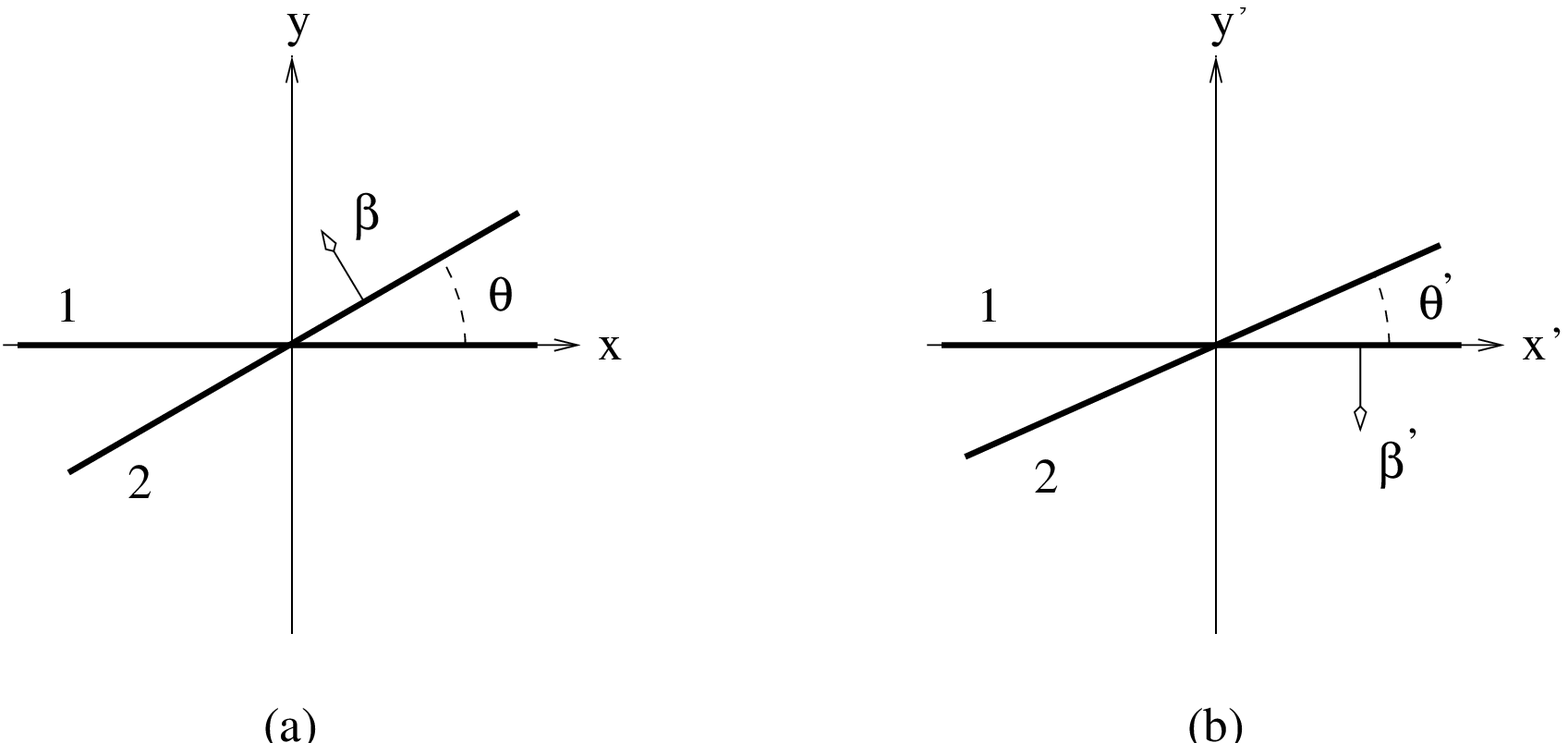}\\
Figure 3: The moving brane intersection from\\
(a) the rest frame of brane 1, (b) the rest frame of brane 2.
\end{center}

\begin{figure}[htp]
\begin{center}
\includegraphics[scale=0.8]{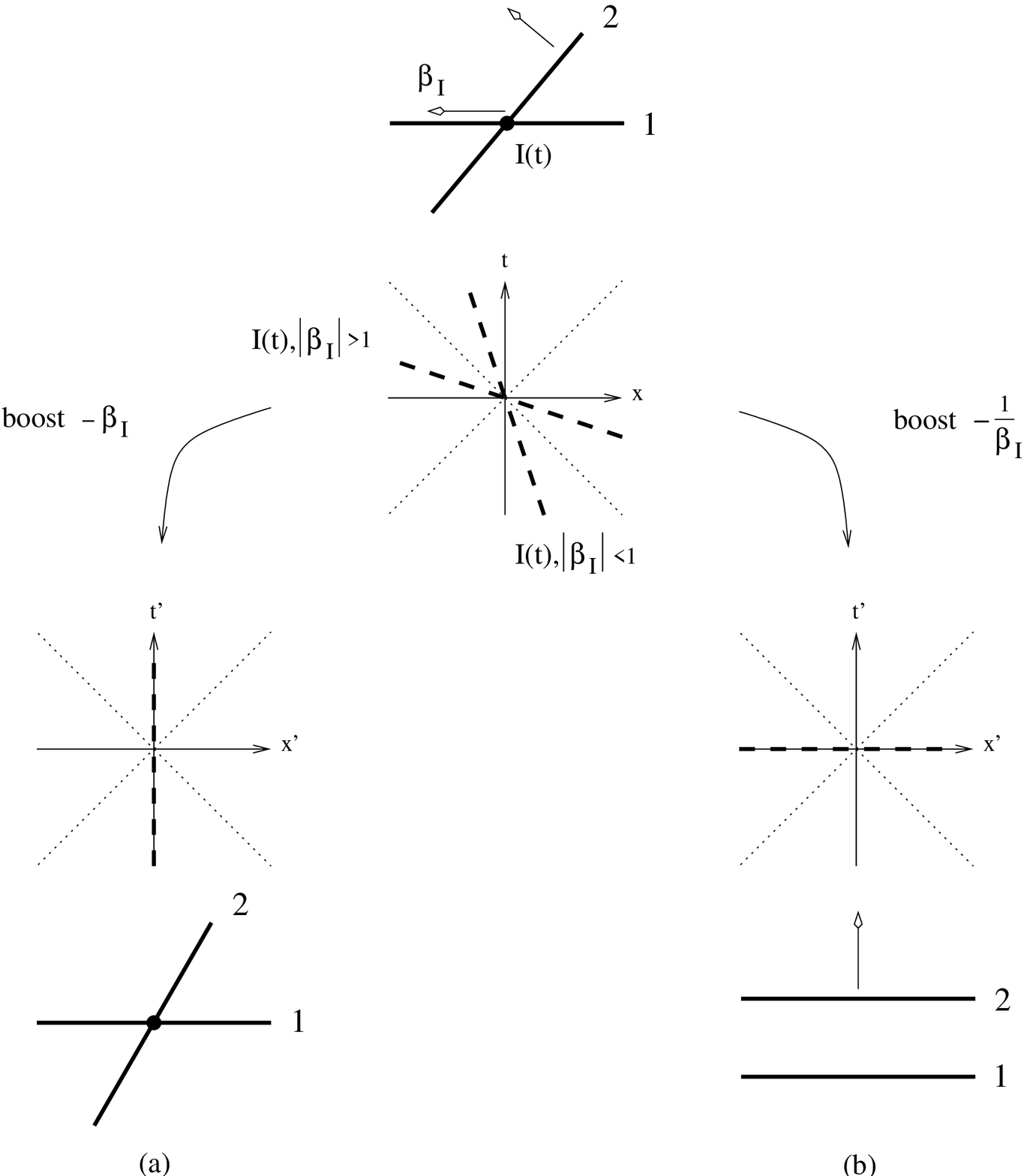}\\
Figure 4: Boosting to the static (a) and parallel (b) configurations
\end{center}
\end{figure}

The next observation becomes useful in describing multiple brane
configurations involving relative motion of the branes. Consider
two branes oriented in the ($x$,$y$) plane as in figure 3(a). The
first is static and lies along the $x$ axis. The second is
oriented at an angle $\theta$ with the $x$ axis and moves with a
velocity $\b$ (which is directed orthogonal to the brane). Now the
apparent velocity of the intersection point\footnote{Of course,
the branes need not intersect if they are displaced in the overall
transverse directions, however, we adopt this nomenclature even in
describing such cases.} along the $x$ axis is
\beq \labell{movie} \bi=-{\b\over\sin\theta}\ , \eeq
which comes from a simple geometric projection of the brane's
motion on the $x$ axis. The interesting point is that while $\b$
is a physical velocity and so have magnitude less than 1, there is
no such bound on this apparent velocity. In fact, $\bi$ reaches
the speed of light for $\sin\theta=\b$ and becomes superluminal
for $\sin\theta<\b$. Note in particular that $|\bi|>1$ may involve
an arbitrarily small physical velocity $\b$ as long as the
intersection angle is correspondingly small.

As may be inferred from the comments at the beginning of the
section, one can easily boost the configuration with $|\bi|<1$ to
a static configuration of intersecting branes by boosting along
the $x$ axis by $-\bi$, \ie one makes a boost along the static
brane. It is amusing then to note that from the point of view of
an observer on the second brane one reaches the static
configuration by boosting along the worldvolume of her brane ---
see figure 3(b). Of course, there is no inconsistency here. In any
frame of reference, one can simply follow the trajectory of the
intersection point
\beq \labell{traject} \vec{x}(t)=\vec{x}_0+\vec{\b}_{\ssc I} t\ ,
\eeq
and then a boosting by $-\vec{\b}_{\ssc I}$ will bring the
intersection to rest, \ie $\vec{x}'(t)=\vec{x}'_0$. Naturally,
this procedure only works when $|\vec{\b}_{\ssc I}|<1$ --- see
figure 4(a).

In the case $|\vec{\b}_{\ssc I}|>1$, the trajectory \reef{traject}
lies outside the lightcone and so one cannot boost to a static
configuration. However, it is possible to boost so that the
intersection becomes a line on a constant time slice. The
necessary operation would be boosting by $-\vec{\b}_{\ssc
I}/|\vec{\b}_{\ssc I}|^2$, \eg in the above example, one would
boost along the $x$ axis by a velocity $-1/\bi=\sin\theta/\b$. In
terms of the full brane configuration, this boost brings one to a
frame where the two D-strings are parallel with one moving with
positive velocity along the $y$ axis. The intersection then occurs
at the instant where this brane crosses the $x'$ axis. This is
shown in figure 4(b).

Of course, the above discussion singles out the case with
$|\vec{\b}_{\ssc I}|=1$. In this case, the trajectory
\reef{traject} lies on the lightcone and it remains null in any
reference frame. We will see that considerations of supersymmetry
select these configurations out as special as well.

\subsection{Enter SUSY}

We would like to understand the constraints imposed by demanding
that the above systems involving branes in motion are
supersymmetric. In section 2.2, we commented on two
time-dependent, yet supersymmetric, configurations of angled
branes related to the static D2-D2 brane pairs. One was
constructed with the M-theory lift and the momentum arose as the
geometrical quantity dual to the non-zero magnetic fields on the
D2-branes, as seen in eq.~\reef{prof}. Of course, re-compactifying
on some transverse direction would reduce this M2-brane
configuration to an identical system of D2-branes. One observation
in eq.~\reef{foreshadow} was that, in this supersymmetric system
of branes in motion, the intersection point moved at the speed of
light. We will see below that this condition appears naturally in
the supersymmetry analysis.

It is simplest to begin by considering the supersymmetry
constraints for the cases with $|\bi|<1$. As described above,
these configurations are easily boosted to a frame where the
intersection is static. Now, the problem of supersymmetric
constraints for static systems of branes intersecting at angles
has received a fair amount of study \cite{BDL,OT,bob1}. While the
details are unimportant, the general result is that supersymmetry
requires the branes to be related by a higher dimensional
rotation, \ie one involving rotations in more than one orthogonal
plane. An example would be an SU(2) rotation acting in an $\IR^4$
subspace of the ten-dimensional Minkowski background. Certainly,
boosting such general configurations will produce configurations
describing a system where the intersection point or surface (if
the branes have common spatial dimensions) moves with a subluminal
velocity. For the simple configurations considered in section 3.1,
the rotation relating the branes is confined to a single plane and
so it would seem that these configurations are necessarily
non-supersymmetric. As a caveat, one should note that the
preceding discussion applies to branes carrying no fluxes. As
described in section 2.2, even for a rotation in a single plane,
one can produce a supersymmetric intersection by introducing an
appropriate electric flux on the branes --- recall
eq.~\reef{stringsol}. Thus, in general, our class of
configurations with $|\bi|<1$ admits supersymmetric solutions if
one is willing to extend the discussion to intersecting D$p$-F1
bound states. The supersymmetric configurations would be boosted
versions of the static configurations discussed at the end of
section 2.2.

For the case with $|\bi|>1$, one can again boost to a canonical
frame as described above, but now this frame corresponds to one
where the branes are parallel but are moving relative to one
another in the transverse space --- see figure 4. This transverse
motion breaks all of the supersymmetries. This simple observation
is supported by the detailed analysis of ref.~\cite{bob2}.

Fortunately, for the analysis of supersymmetry in the case
$|\bi|=1$, we can borrow from the discussion of M5-branes given in
ref.~\cite{bob2}. As usual the supersymmetry constraints are
determined from considering the compatibility of the supersymmetry
projections for the individual branes, $\G_i\eps=\eps$, analogous
to the discussion in section 2.1. As for the general case of
intersecting branes \cite{BDL}, the projections may be written as
\beq \labell{block}
\G_2=S\,\G_1\,S^{-1}\ ,
\eeq
where $S$ is the Lorentz transformation that relates the first
brane to the second. The essential observation \cite{bob2} is that
for the case where $|\bi|=1$ the necessary transformation can be
characterized as a `null rotation' --- see, for example,
\cite{penrose}. For example, to produce the configuration
illustrated in figure 3(a), the relevant transformation is
\beq \labell{nullrot}
S(\beta)=\exp\left[\,\ga\b\,(\Sigma_{ty}-\Sigma_{xy})\right]\ , \eeq
where $\Sigma_{\mu\nu}$ denote the generator of a boost or
rotation in the $(x^\mu,x^\nu)$ plane. It is a straightforward
exercise to show that the relevant `rotation' parameter is
$\gamma\b$ where $\beta$ is the magnitude of the brane's velocity
appearing in the figure (and as usual $\ga=1/\sqrt{1-\b^2}$). The
defining characteristic of a null rotation is that is leaves a
null vector invariant, \ie $(1,-1,0,\cdots)$ in the case of
eq.~\reef{nullrot}. Hence, beginning with coincident branes, after
applying a null rotation to one of them, this null vector remains
common to the tangent space of both worldvolumes. That is, the
intersection point or surface moves along this null direction.

Following the analysis of ref.~\cite{bob2}, one uses the property
that
\beq\labell{prper} \G_1\,S^{-1}=S\,\G_1 \eeq
to show that compatibility of the two supersymmetry projections
requires $S^2\eps=\eps$. Further, for the spinor representation,
the Lorentz generators are simply $\Sigma_{\mu\nu}={1\over2}
\G_{\mu\nu}$, from which one finds that the square (and higher
powers) of the generator appearing in the null rotation vanishes.
Hence, the previous compatibility condition reduces to a simple
projection along the invariant null direction. That is, for
eq.~\reef{nullrot}, one has
\beq \labell{nullpro} \left(\G_t-\G_x\right)\eps=0\ .\eeq
Such a null projection is familiar, \eg for a gravity wave
\cite{susywave} or massless particle propagating along the $x$
axis. It follows from the construction that this projection
operator commutes with the original projection operator $\G_1$.
Hence these two projections are compatible and the brane system
with a null intersection is one-quarter BPS.

Of course, this general result may have been anticipated because
the null intersections arose in section 2.2 as a dual description
of a one-quarter BPS configuration of parallel branes and
antibranes. For example, the M-theory lift of the D2-D2 pair
described by eq.~\reef{specsol} gave a pair of M2-branes with a
null intersection. As a check, it is interesting to verify that
the same supersymmetry conditions arise in the original
configuration. For specificity consider the choice of fields $E_2=B_2=-\veps$
and $E_1=B_1=0$. Now, applying the preceding discussion to the
corresponding lift to a pair of M2-branes with a null
intersection, one finds:\footnote{Note that here $y$ specifies a
common worldvolume direction orthogonal to the hypersurface where
the null rotation acts. Here and in section 2.2, the coordinate
$z$ plays the role of $y$ in the previous paragraph.}
$\G_{txy}\eps=\eps$ and $(\G_t-\G_x)\eps=0$. Working directly with
the D2-branes, one finds that the worldvolume fluxes on the second
brane are consistent with the choice $E_1=E_1^-$ in
eq.~\reef{parsol}. Then the compatibility condition in
eq.~\reef{fullsol} becomes $\G_y\eps=-\eps$ while the inequality
is automatically satisfied. The original supersymmetry constraint
associated with the D2-brane with vanishing fluxes is simply
$\G_{txy}\eps=\eps$. Combining these two projections yields
$\G_{tx}\eps=-\eps$, from which the desired null projection
readily follows.

\subsection{Worldvolume Picture}

Next we investigate excitations of these systems involving branes
in motion. The approach followed in this section is to study the
worldvolume theory. For simplicity, we consider a symmetric
configuration of D-strings as illustrated in figure 5. To leading
order, the low energy worldvolume theory on the D-strings is the
reduction of ten-dimensional U$(2)$ super-Yang-Mills theory to two
dimensions. The bosonic Lagrangian, which is sufficient for our
purposes, is
\beq \cL={2\pi\ls^2\over g}\tr \left( -\frac{1}{4}
F_{ab}F^{ab}-\frac{1}{2}D_a \Phi^i D^a \Phi^i +
\frac{1}{4}[\Phi^i,\Phi^j][\Phi^i,\Phi^j]\right)\ .
\labell{SYMlag} \eeq
Here, $F_{ab}$ is the non-Abelian U$(2)$ field strength and
$\Phi^i$, with $i=2,\ldots,9$, are the eight adjoint scalars. Our
conventions are such that $\Phi^i=(\Phi^i)^{(n)}\tau_n$, where the
Hermitean generators of U$(2)$ are chosen to be: $\tau_0$, the
two-by-two identity matrix, and $\tau_{1,2,3}$, the standard Pauli
matrices. For simplicity, we work in static gauge where the
timelike (spacelike) coordinate on the worldvolume matches the
background coordinate $t$ ($x$).

The configuration shown in figure 5 is studied by introducing a
background vacuum expectation value (vev) to one of the scalars,
$\Phi^2$, say, which represents displacements of the worldvolume in
the $y$ direction. The profile of the two strings in this
symmetric configuration is $y=\pm(x\sin\theta+t\b)/\cos\theta$, up to a
constant shift in $t$ or $x$. This then corresponds to the
background scalar vev
\beq\labell{vev} \Phi_0^2\equiv V\tau_3=
\frac{x\sin\theta+t\beta}{2\pi\ls^2\cos\theta}\tau^3\ . \eeq
Note that this expectation value (and all other fields zero) is a
solution of the equations of motion following from
eq.~\reef{SYMlag}. One could also displace the strings along
another orthogonal direction by adding an additional constant vev,
\eg $\Phi_0^3=d\,\tau_3$, so that the D-strings do not actually
intersect. This additional complication does not qualitatively
modify the results below. Also note that in this symmetric
configuration, the intersection between the D-strings lies on the
$x$ axis and so its apparent velocity is still given by
eq.~\reef{movie}. Now, the low energy theory \reef{SYMlag} will
give an adequate description of the physics when the fields are
slowly varying. However, we should comment that even though the
backgrounds of interest may describe intersections with an
apparent velocity at or near the speed of light, this does not
contradict the requirement of slowly varying fields. The latter
will all ways be satisfied if $\b$ and $\sin\theta$ are
sufficiently small, but the latter does not restrict their ratio,
which determines $\bi$.

\begin{center}
\includegraphics[scale=0.6]{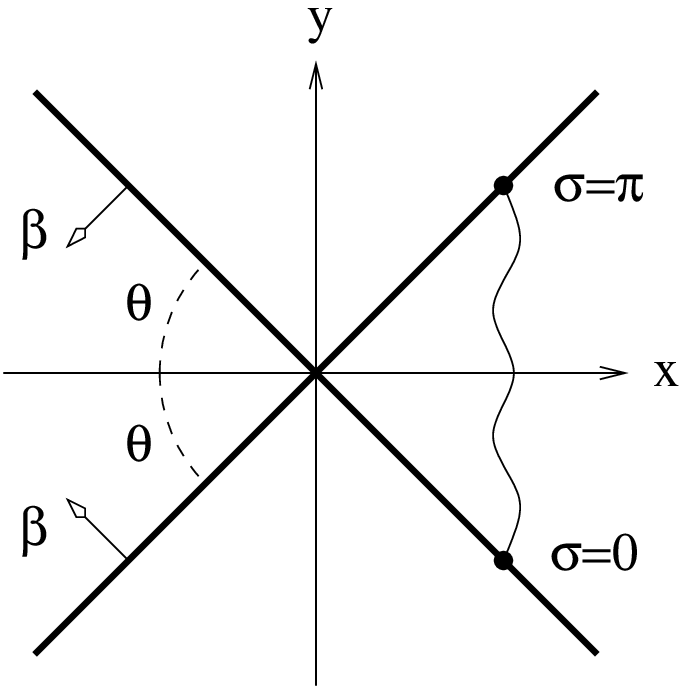}\\
Figure 5: A symmetric D-brane configuration, with attached string
\end{center}

With this background, gauge symmetry is spontaneously broken to
U$(1)\,\times\,$U$(1)$. The unbroken U$(1)$'s are associated with
$\tau_{0,3}$ components and the corresponding fields correspond to
string modes which move along one string or the other. We are most
interested in the $\tau_{1,2}$ components of the fields as they
correspond to strings connecting one D-string to the other. The
simplest case\footnote{The scalar $\Phi^2$ mixes with the gauge
field, but shows qualitatively similar behavior to that found
below.} to consider are the transverse scalars orthogonal to the
plane of the motion, \ie $\Phi^i$ with $i=3,\ldots,9$. We denote
the corresponding off-diagonal components as the complex scalars
$\phi^i$. When we expand around the background \reef{vev}, the
linearized equation of motion of each field is simply
\beq \labell{linex} -\Box\phi+2V^2\phi=0\ , \eeq
with what appears to be a spacetime dependent mass term. Now, the
precise behavior of the solutions depends on whether the linear
combination of $x$ and $t$ entering $V^2$ is spacelike, timelike
or null --- which corresponds to an apparent velocity $\bi$ which
is subluminal, superluminal or null. Hence we consider each of
these cases in turn:

{\bf i) Subluminal ($\b<\sin\theta$):} Here the first step is to
make the coordinate transformation
\beq\labell{bonk}
u=t\sin\theta+x\b\ , \qquad
w=t\b+x\sin\theta\ .
\eeq
Note that under this transformation, the line element on the
D-string worldvolume becomes
\beq
\labell{linel}
ds^2=-dt^2+dx^2={1\over\sin^2\theta-\b^2}\left(-du^2+dw^2\right)\ ,
\eeq
and so we may think of $u$ ($w$) as a time (space) coordinate. In
these coordinates, the equation of motion \reef{linex} becomes
\beq\labell{psieq}
\left(\pf_u^2-\pf^2_w\right)\phi+U(w)\phi=0\ ,
\eeq
where
\beq\labell{potentate}
U(w)=\frac{1}{2\pi^2\ls^4(\sin^2\theta-\b^2)\cos^2\theta}\,w^2\ .
\eeq
Making a standard separation of variables, $\phi=e^{-i\Omega u}\psi(w)$,
the equation of interest reduces to
\beqa\labell{PP}
-\pf_w^2\psi +U(w)\psi=\Omega^2\psi\ .
\eeqa
Now eq.~\reef{PP} has the form of the time-independent
Schr\"odinger equation for a simple harmonic oscillator potential.
Comparing to eq.~\reef{potentate}, the oscillator frequency and
the energy eigenvalues are:
\beq \labell{fix}
\omega=\frac{1}{\pi\ls^2\cos\theta\sqrt{2(\sin^2\theta-\beta^2)}}\ ,
\qquad E=2\Omega^2=\left(n+{1\over2}\right)\omega\ ; \quad
n=0,1,2,\ldots\ . \eeq
Further we can write out the precise solutions (see, \eg
ref.~\cite{quant}) but it suffices to say that they are localized
around the origin $w=0$.

Now we observe that the transformation \reef{bonk} essentially
amounts to the Lorentz transformation that brings the branes to a
static configuration, as described at the end of section 3.1. In
the original frame $w=0$ corresponds to the location of the
intersection point between the two D-strings. The localization
found above indicates that for the `massless' string modes
connecting the two D-strings, their excitations are carried along
by the motion of the intersection in this the subluminal case.
This is analogous to the results found in ref.~\cite{BDL} for the
case of a static brane intersection. Note that for lowest-lying
modes, the modes are very well confined to the vicinity of the
intersection and so the higher order (cubic and quartic)
interactions in the Lagrangian \reef{SYMlag} should be negligible.
Hence the linearized equations of motion \reef{linex} studied here
should give an accurate description of the physics.

{\bf ii) Superluminal ($\b>\sin\theta$):} In this case, the
analysis is essentially the same as above. However from
eq.~\reef{linel} we see that the role of the new coordinates is
reversed, \ie $u$ is space coordinate while $w$ is the time
coordinate. Further since $(\sin^2\theta-\b^2)$ has changed sign
in this regime, the potential \reef{potentate} in the effective
Schr\"odinger equation \reef{PP} is now an {\it inverted} harmonic
oscillator. Exact analytic solutions can be still written down in
terms of parabolic cylinder functions (see, \eg ref.~\cite{whit}) and
in accord with one's intuition, the wavefunctions are not
localized near the origin.

The physical interpretation of these results is as follows. In
this case, the coordinate transformation \reef{bonk} is
essentially the boost which puts the D-strings in the canonical
frame where they are parallel but move relative to one another
along the $y$ axis. The failure of the wavefunctions $\psi(w)$ to
be localized reflects the fact that these excitations cannot keep
up with the intersection point. That is, if we construct an
initial wave packet that is localized near the intersection, say,
$x=0$ at $t=0$, this excitation falls behind the intersection
which moves off to the left along $x=-|\bi|\,t$. Thus the wave
packet will eventually enter a regime where the expectation values
of the background scalars are large and so the low energy analysis
becomes unreliable as the relevant mass scales are string scale.

As an aside, it is interesting to remark that if one creates a
wavepacket with both $\tau_{1,2}$ components in more than one of
the transverse scalars, then this excitation would generate a
nontrivial coupling to the RR four-form potential
\cite{core,dielectric}. Hence such an excitation would introduce a
nontrivial D3-brane component, which presumably gets stretched out
between the D-strings.

{\bf iii) Null ($\b=\sin\theta$):} In this case, the coordinate
transformation \reef{bonk} becomes degenerate and so we must
modify the analysis somewhat. It is useful to define light-cone
coordinates
\beq \labell{nullcoor}
x^\pm={t\pm x\over\sqrt{2}}\ ,
\eeq
in terms of which the equation of motion \reef{linex} becomes
\beq\labell{conic}
\pf_-\pf_+\phi+\frac{\tan^2\theta}{2\pi^2\ls^4}{x^+}^2\phi=0\ .
\eeq
We can readily solve this equation with the usual separation of
variables but it is useful to make yet another change of
coordinates: $u={x^+}^3/\ls^2$. Note that with these new
coordinates, the worldvolume metric becomes
\beq \labell{linel2} ds^2=-dt^2+dx^2=-dx^+dx^-=
\frac{1}{3}\left(\frac{\ls}{u}\right)^{2/3}
\left(-du\,dx^-\right)\ , \eeq
while the equation of motion reduces to
\beq\labell{conic2}
\pf_-\pf_u\phi+\frac{\tan^2\theta}{6\pi^2\ls^2}\phi=0\ . \eeq
The latter takes the form of an ordinary Klein-Gordon equation for
a scalar field with mass $M=\tan\theta/\sqrt{6}\pi\ls$. Hence we
can easily write down solutions as
\beq\labell{nusol} \phi=e^{ip_-x^-}e^{iM^2u/p_-}
=e^{ip_-x^-}e^{iM^2{x^+}^3/\ls^2p_-}\ . \eeq

In these coordinates, the intersection lies at $x^+=0$, and the
obvious interpretation of $p_-$ is as the momentum of a plane wave
along the null direction $x^-$. The explicit form of the wave
functions \reef{nusol} does not immediately suggest the proper
physical interpretation for these results. The introduction of $u$
is useful for these purposes: Begin by considering the equation of
motion \reef{conic2}. Regarding this as a massive scalar wave
equation in flat Minkowski space, it is natural to construct wave
packets which are peaked near timelike geodesics
--- see figure 6. As the flat metric $d\tilde{s}^2=-du\,dx^-$ is
related to the worldvolume metric \reef{linel2} by a conformal
transformation, the causal nature of this propagation should
survive for these excitations of the D-strings. That is, that wave
packets naturally follow timelike trajectories. Using the
coordinate transformation $u={x^+}^3/\ls^2$, we see that a typical
trajectory might be $x^-={x^+}^3/\ls^2$, which is easily verified
to be timelike --- see figure 6. Hence we see that as in the
subluminal case, the excitations are `swept' along by the null
intersection. However, as in the superluminal case, the wave
packets can not keep up with the intersection. Hence in this case,
the excitation of the fields may not grow dramatically but
invariably a wavepacket enters a regime where the expectation
values of the background scalars are large and so the low energy
analysis again becomes unreliable.

\begin{center}
\includegraphics[scale=0.5]{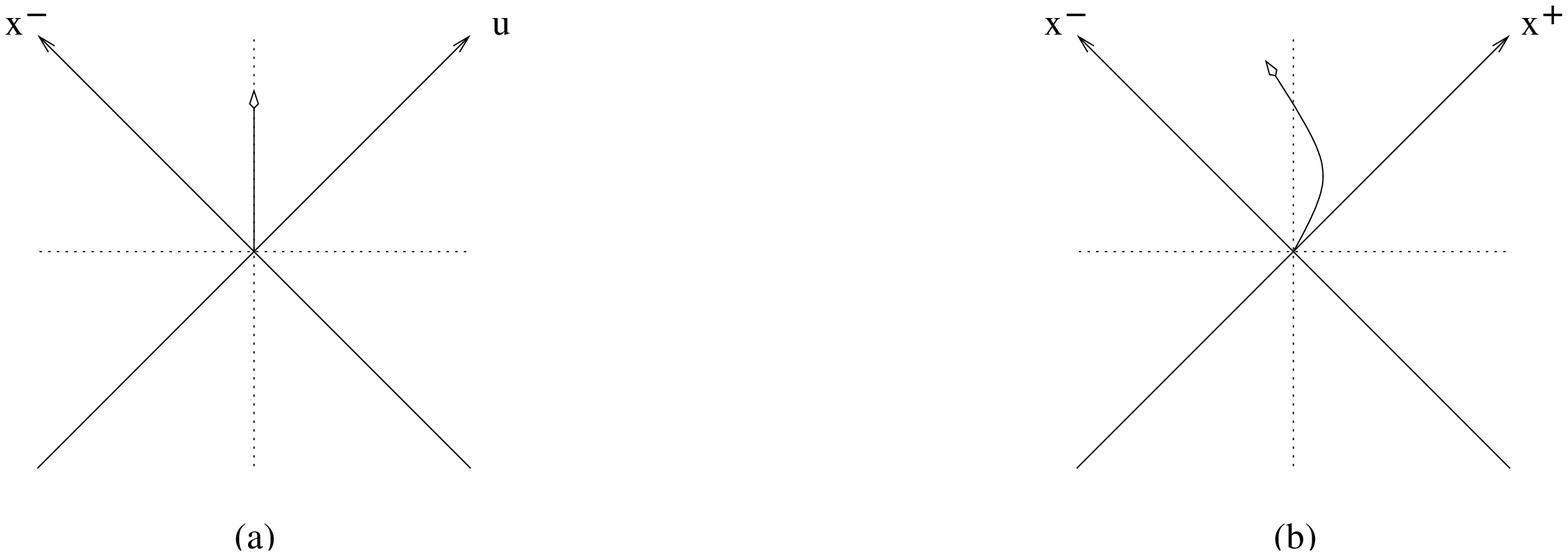}\\
Figure 6: a) Typical timelike geodesic in the ($u,x^-$) plane\\
and b) the corresponding trajectory in the ($x^+,x^-$) plane
\end{center}

\subsection{Macroscopic Strings}

One can also investigate the string modes connecting the two
branes in motion by examining directly the worldsheet equations of
the fundamental strings. One can naturally quantize these modes
and calculate the perturbative string spectrum. However, for the
purposes of gaining some physical insight into the behavior of
these systems, we will satisfy ourselves here with considering
particular solutions describing the behaviour of macroscopic
strings connecting the two branes. In this analysis we again
consider the symmetric configuration of moving D-strings examined
in the last section --- see figure 5.

Information about the motion of the D-strings is encoded in the
boundary conditions on the worldsheet fields. Those boundary
conditions for the configuration under consideration are
\beqa
\pf_\sigma[T+\b(Y\cos\theta + X\sin\theta)]_{\sigma=0}=0\ ,
&\quad & \pf_\sigma[T-\b(Y\cos\theta -X\sin\theta)]_{\sigma=\pi}=0\ ,\\
\quad \pf_\sigma[X\cos\theta- Y\sin\theta]_{\sigma=0}=0\ ,
&\quad & \pf_\sigma[X\cos\theta+Y\sin\theta]_{\sigma=\pi}=0\ ,\\
\quad [T\b+(Y\cos\theta+ X\sin\theta)]_{\sigma=0}=0\ , &\quad &
[T\b-(Y\cos\theta-X\sin\theta)]_{\sigma=\pi}=0\ .
\eeqa
The fields $\{X^i\}_{i=3\ldots 9}$, that map the worldsheet into
the spatial  directions transverse to the plane of the
configuration, have standard Dirichlet boundary conditions, and
associated open string mode expansions --- see, \eg
ref.~\cite{clifford}.

When the intersection moves slower than the speed of light
($\sin\theta>\b$), some representative solutions  of $\Box X^a=0$
may be written
\beqa T(\sigma,\tau) &=& 2\ls^2 p \tau - \frac{\b}{\cos\theta}\ls
H(\tau)I(\sigma)\ , \qquad
Y(\sigma,\tau) = \ls H(\tau)J(\sigma)\ ,\nonumber\\
X(\sigma,\tau) &=& -\frac{2\b}{\sin\theta}\ls^2 p \tau +
\frac{\sin\theta}{\cos\theta}\ls H(\tau)I(\sigma)\ .\labell{slowexp2}
\eeqa
The functions $H$, $I$ and $J$ and given by
\beqa
H(\tau)&=&A\cos\delta\tau+B\sin\delta\tau\ ,\nonumber\\
I(\sigma))&=&\sin\delta\sigma-\frac{\cos\theta}{\sqrt{\sin^2\theta-\b^2}}\cos\delta\sigma\ ,\labell{It}\\
J(\sigma)&=&\sin\delta\sigma+\frac{\sqrt{\sin^2\theta-\b^2}}{\cos\theta}\cos\delta\sigma\
,\nonumber \eeqa
where $\delta$ is given by
\beq\labell{slowdel}
\tan\pi\del=\frac{2\cos\theta\sqrt{\sin^2\theta-\b^2}}{\sin^2\theta-\cos^2\theta-\b^2}\ .
\eeq
$A$ and $B$ are real constants, corresponding to the simple choice
of oscillators that we have excited.

\begin{center}
\includegraphics[scale=0.3]{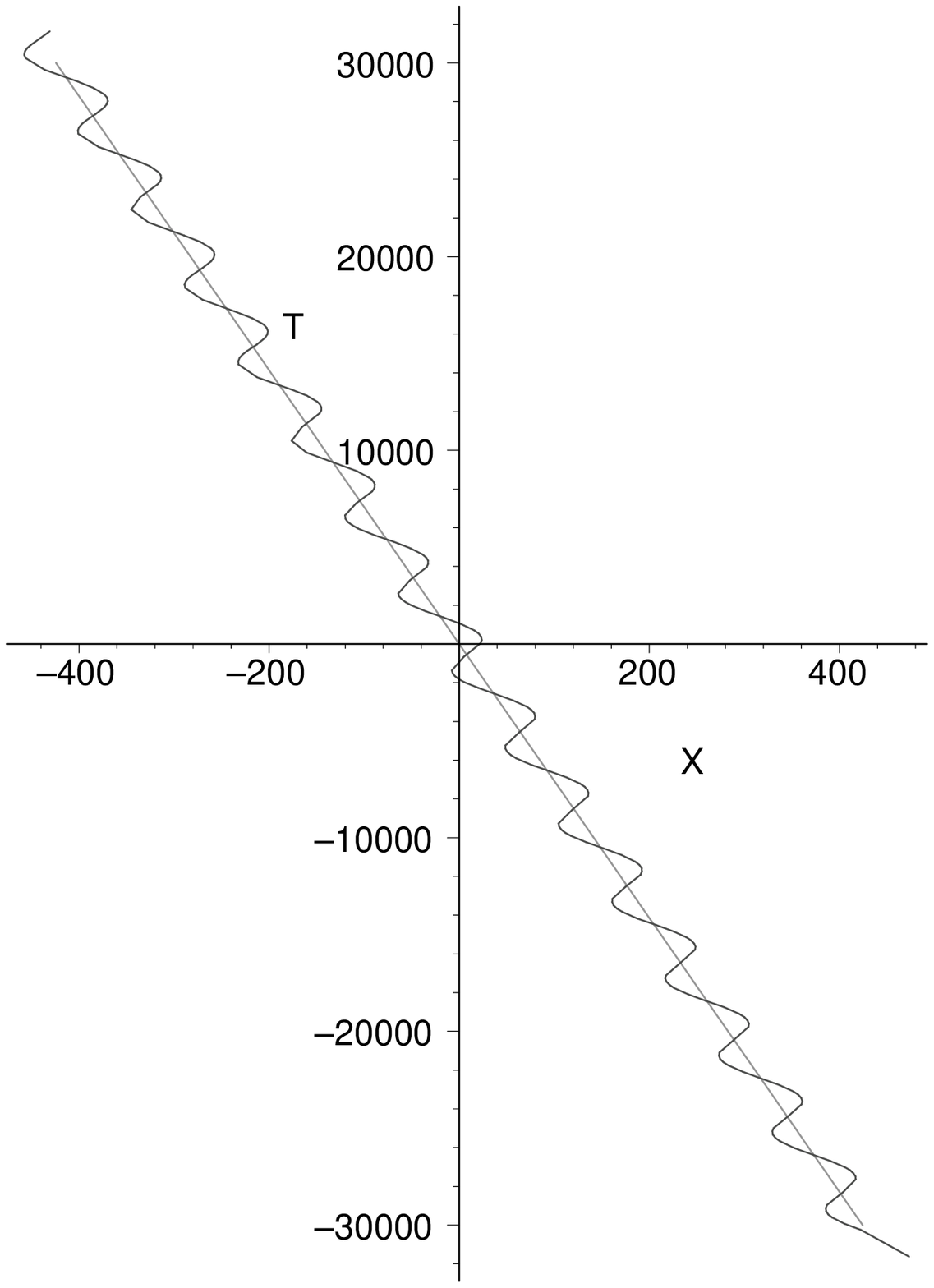}\\
Figure 7: Oscillatory motion of the string\\
about the moving intersection, for $\b<\sin\theta$
\end{center}

In figure 7 we display a plot of the trajectory, along the $x$
axis, of one of the endpoints of the string, calculated from
eqs.~\reef{slowexp2}. One clearly sees the oscillatory motion of
the string around the moving intersection (in this direction, the
two endpoints move in phase with one another). The motion of the
endpoints in the $y$ direction, perpendicular to the motion of the
intersection, is also oscillatory, but now the two endpoints are
in anti-phase, indicating that they pass over one another when
they cross the intersection. Of course, since $\b<\sin\theta$,
this configuration could be boosted to a standstill, in which case
the behaviour just described is exactly what one would expect ---
the confinement of modes to the region around the intersection.
This localisation matches the behavior found for the microscopic
strings in the U$(2)$ gauge theory in section 3.3.

Note that in the general solution for the worldsheet modes, one
finds modes with non-integer oscillator frequencies. The
expression $\delta$ in eq.~\reef{slowdel} is the amount by  which
the frequencies are shifted from integer. Notice that $\delta$
vanishes in the limit $\b\to\sin\theta$, signalling the
re-appearance of integer moded oscillators in the supersymmetric
case. However, it appears that, in the same limit, the $X$ and $T$
mode expansions become singular. This behaviour was noticed in the
mode expansions of strings joining the D2-$\dir{\textrm{D2}}$ pair
of ref.~\cite{strings}, where it is an effect of the worldvolume
gauge fields. There it was argued that the worldsheet CFT actually
remains non-singular, so that the spacetime supersymmetric theory
is well-defined.

If we continue to the regime where the intersection velocity is
superluminal ($\sin\theta<\b$), the phase shift $\delta$ becomes
imaginary.  The analogous solutions to those in
eq.~\reef{slowexp2} are
\beqa
T(\sigma,\tau) &=& 2\ls^2 p \tau -
\frac{\b}{\cos\theta}\ls\bar{H}(\tau)\bar{I}(\sigma)\ , \qquad
Y(\sigma,\tau) = -\ls \bar{H}(\tau)\bar{J}(\sigma)\ ,\nonumber\\
X(\sigma,\tau) &=& -\frac{2\b}{\sin\theta}\ls^2 p \tau -
\frac{\sin\theta}{\cos\theta} \ls \bar{H}(\tau)\bar{I}(\sigma)\ ,
\labell{fastexp}
\eeqa
where, now,
\beqa
\bar{H}(\tau)&=&\bar{A}\cosh\bar{\delta}\tau+\bar{B}\sinh\bar{\delta}\tau\ ,\nonumber\\
\bar{I}(\sigma)&=&\sinh\bar{\delta}\sigma+
\frac{\cos\theta}{\sqrt{\b^2-\sin^2\theta}}\cosh\bar{\delta}\sigma\ ,\labell{fastfunk}\\
\bar{J}(\sigma)&=&\sinh\bar{\delta}\sigma+
\frac{\sqrt{\b^2-\sin^2\theta}}{\cos\theta}\cosh\bar{\delta}\sigma\
,\nonumber \eeqa
and
\beq\labell{fastdel}
\tanh\bar{\delta}\pi=
\frac{2\cos\theta\sqrt{\b^2-\sin^2\theta}}{\sin^2\theta-\cos^2\theta-\b^2}\
. \eeq
As we have seen, in this case it is not possible to boost the
intersection to a standstill. Rather, as we know the crossed
branes are now equivalent to parallel branes moving away from one
another, as in figure 4(b), we expect to find strings that grow
without bound in time. The expressions above indeed exhibit this
behaviour, implicit in the hyperbolic functions that have replaced
the sines and cosines of eqs.~\reef{It}. Therefore at large times,
these functions dominate the evolution of these strings. In
particular, one finds that the horizontal velocity of the string
is essentially constant with $dX/dT(\sigma)\simeq\sin\theta/\b$.
Of course, this precisely matches the behavior expected from the
discussion at the end of section 3.1. Hence, the string lags
behind the intersection, while being stretched apart in the $y$
direction. The gauge theory also displayed a version of this
macroscopic physics, as we saw in section 3.3.

\section{Discussion}
\label{disc}

In this paper, we considered two topics in the physics of
D-branes. The first was generalized supersymmetry conditions for
parallel D-$\dir{\textrm{D}}$ pairs and the second was brane
intersections in motion. The connection between these
configurations was that in certain situations, they naturally
arose as dual descriptions of the same system.

Supertubes are a fascinating arena in which to study D-brane
physics \cite{stube}--\cite{lug},\cite{dangle}. They seem to
provide a counter-example to several aspects of `popular string
lore,' one example being that brane-antibrane systems are
necessarily nonsupersymmetric.  In the case of parallel
D-$\dir{\textrm{D}}$ pairs, we found generalized flux
configurations for two-branes which allowed the systems to be
one-quarter BPS. Previous studies, which produced extensions of
supertube configurations, generally found the relevant electric
field took the `critical' value given in eq.~\reef{crite}.
However, one of the outcomes of our analysis was that a critical
electric field is not a crucial ingredient to producing a
supersymmetric system. We should note that a similar conclusion is
implicit in ref.~\cite{lug} which studied D3-$\dir{\textrm{D3}}$
and D4-$\dir{\textrm{D4}}$ pairs. It is true that here we focussed only
on planar configurations while the supertube allows for more
interesting geometries. However, even for a supertube, one could
shift the electric field away from its critical value by boosting
along the axis of the tube. This boost would create an additional
component to the electric field orthogonal to the axis,
corresponding to introducing fundamental strings which wind around
the tube. In any event, an interesting extension of our work would
be to consider in detail how the tachyonic modes in the
perturbative string spectrum of D-$\dir{\textrm{D}}$ pair are
lifted as the background fluxes approach their supersymmetric
values.

Lifting a supersymmetric D2-$\dir{\textrm{D2}}$ pair to
eleven-dimensional M-theory produces a BPS configuration of
M2-branes at an angle in which the intersection line moves at
lightspeed. In considering general configurations of intersecting
D-branes in motion, one finds that supersymmetry naturally selects
out cases where the intersection is null. Related M5-brane
configurations were previously considered in ref.~\cite{bob2}.
Those authors provided a beautiful construction which can be
applied quite generally to construct branes with null
intersections by the application of a null rotation. The
compatibility of the supersymmetry of the branes involves a null
projection \reef{nullpro} along the tangent to the trajectory of
the intersection. Similar null projections appear for related
systems of branes such as superhelices\cite{shelix,scurve},
M-ribbons\cite{ribm} or giant gravitons\cite{gravitons}.

Further analysis of excitations of the intersecting D-branes in
motion gave interesting results. In the case where the
intersection point moves at less than the speed of light, the
excitations involving strings connecting the two branes are simply
swept along by the motion of the intersection. This result
appeared both in the investigations of the low energy worldvolume
theory and of macroscopic strings. In the case of superluminal
motion of the intersection, both approaches revealed
instabilities. The macroscopic strings were unable to keep up with
the intersection point and so were continually stretched as they
fell behind --- a result that is not surprising given the
description of these intersections in a canonical frame with
parallel branes moving apart. Similarly wave packets in the low
energy theory would fall behind the superluminal intersection
point. Analysis of the null intersections revealed a similar
falling behind and stretching instability. We note that the
stretching instability of the macroscopic strings need not be
considered as pathological as, say, the runaway behavior of a
tachyonic mode. Similar stretching instabilities have recently
been discussed for closed strings in plane wave backgrounds
\cite{stretch}. The amplitude of the low energy excitations do not
seem to show a tachyonic or runaway instability in that their
amplitude does not diverge as they evolve in time. Rather within
the low energy analysis, the relevant solutions of eq.~\reef{PP}
have the asymptotic behavior $|\psi|^2\sim 1/w$ in general.

Recently there has been a great deal of interest in time dependent
backgrounds in string theory \cite{tyme,null,sic}. One could study
the systems studied above as simple time dependent backgrounds for
open strings. In many respects, the null intersections are closely
related to null orbifolds \cite{null}. In particular, the
construction of both is based on a null rotation. Although the
null orbifolds are supersymmetric, one finds that in many cases
these backgrounds are `fragile' in that probes generically produce
regions of strong curvature \cite{sic}. This behavior might be
regarded as analogous to the stretching instability of string
modes as they fall behind the intersection point. Of course for
the null orbifolds, the `probe cataclysms' can be suppressed by
including a spatial translation in the orbifold group \cite{sic}.
In the present case, similarly the excitation of the interbrane
strings could be suppressed by increasing the threshold with an
additional separation of the D-branes in the transverse space.
Note, however, that the gravitational effects in the orbifold case
are created independent of the value of the string coupling
constant. In the present case, it seems that these string modes
would only be created perturbatively when an intersection collides
with modes travelling down one of the individual branes. Hence
their appearance can be suppressed by reducing the string
coupling. In any event, these null intersections are worthy of
further study for the potential insight that they may give for the
problem of time dependent backgrounds.


\section*{Acknowledgements}
We would like to acknowledge useful conversations with
Martin Kruczenski, David Mateos, David Page and Konstantin
Savvidy. DJW and RCM were supported in part by NSERC of Canada and
Fonds FCAR du Qu\'ebec. DJW gratefully acknowledges the ongoing
hospitality of the Physics Department at the University of
Waterloo during this research. While this paper was being
completed, ref.~\cite{scissors} appeared which overlaps with the
material in section~\ref{lorentz}.

\appendix
\section{T-duality transformations}
\label{hoho}

In general, the effect of T-duality parallel to a flat D2-D0-F1
bound state is to produce a D1-F1 bound state moving, at an angle
to the T-duality direction, in the plane of the original brane
system. The presence of the lower-dimensional constituents of each
bound state induces a BI field strength on each D-brane. The
momentum and the angle are understood to arise from the
reinterpretation of certain components of the D2-brane gauge
potential as new scalars transverse to the D1-brane's worldvolume.
If we T-dualise in a direction $x^p$ along the D2-brane this
relationship can be written, in terms of the field strength, as
--- see, \eg ref.~\cite{dielectric}:
\beq\labell{dict}
F_{ap}\longrightarrow \pf_a \Phi^p\ ,
\eeq
 where the transverse
scalar is related to the transverse coordinate by
$\Phi^p=x^p/(2\pi\ls^2)$, and determines the profile of the
D-string. This is sufficient to calculate the angle and momentum
produced by T-duality, but is not the whole story, as it tells us
nothing about the electric flux on the D-string. The simplest way
to calculate this is to infer the result from the action of the
T-duality on the worldsheet boundary conditions for the open
strings ending on the D-string or the D2-brane. The details of
this procedure can be found in, \eg ref.~\cite{clifford,more}, and
give values for the angle and momentum that agree with those
determined from eq.~\reef{dict}. We start with the D-string
configured as in figure 8, with an electric field denoted by $e$.
Then the T-dual configuration is a D2-brane filling the $(x,y)$
plane, with BI field strength
\beq F=\tilde{E}\df x\wedge\df t+E\df y\wedge\df t+B\df y\wedge\df x\ ,
\label{genFagain}
\eeq
where, if the T-duality was made along the $x$ axis,
\beq\labell{Tx} 
\tilde{E}=\frac{\beta}{2\pi\ls^2\sin\theta}\ ,
\quad E=\frac{e}{\gam\sin\theta}\ , \quad B=\frac{\cot\theta}{2\pi\ls^2}\ .
\eeq
Alternatively, T-duality along the $y$ axis gives
\beq\labell{Ty}
\tilde{E}=\frac{e}{\gam\cos\theta}\ , \quad
E=-\frac{\beta}{2\pi\ls^2\cos\theta}\ , \quad  B=-\frac{\tan\theta}{2\pi\ls^2}\ .
\eeq
Note that $\gam=1/\sqrt{1-\b^2}$, as usual.

\begin{center}
\includegraphics[scale=0.6]{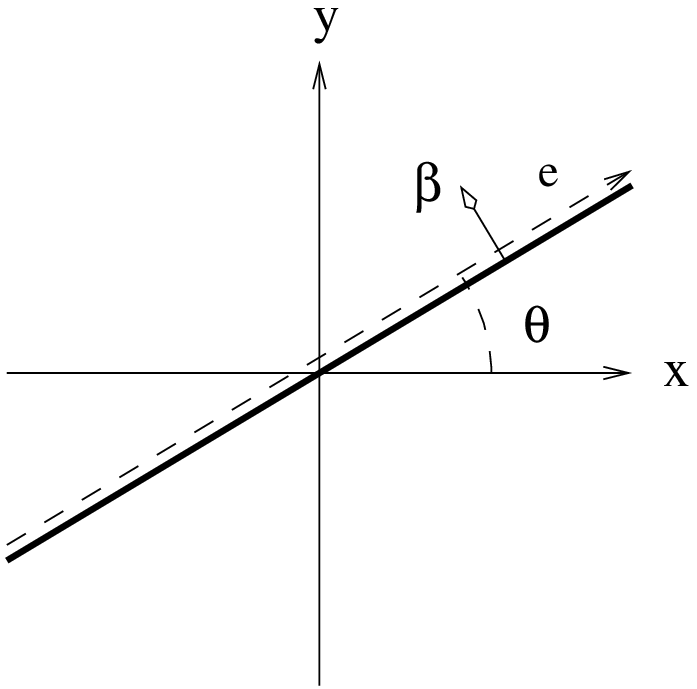}\\
Figure 8: The dual D-string with flux
\end{center}

\end{document}